\documentclass[prd,a4paper,aps,showpacs,nofootinbib,nobibnotes,superscriptaddress]{revtex4}

\begin{document}

\newcommand{\pkt}{\; .}
\newcommand{\kma}{\; ,}
\title{CMB anisotropies due to cosmological magnetosonic waves}

\date{\today}
\author{Tina Kahniashvili}
\email{tinatin@phys.ksu.edu} \affiliation{Department of Physics,
Kansas State University, 116 Cardwell Hall, Manhattan, KS 66506,
USA} \affiliation{CCPP, Department of Physics, New York
University, 4 Washington Place, NY 10003, USA
}\affiliation{National Abastumani Astrophysical Observatory, 2A
Kazbegi Ave, GE-0160 Tbilisi, Georgia}
\author{Bharat Ratra}
\email{ratra@phys.ksu.edu}
\affiliation{Department of Physics, Kansas State University,
116 Cardwell Hall, Manhattan, KS 66506, USA}

\date{November 2006 \hspace{0.3truecm} KSUPT-06/6}

\begin{abstract}
We study scalar mode perturbations (magnetosonic waves) induced
by a helical stochastic cosmological magnetic field  and derive
analytically the corresponding cosmic microwave background (CMB)
temperature and polarization anisotropy angular power spectra. We
show that the presence of a stochastic magnetic field, or an
homogeneous magnetic field, influences the acoustic oscillation
pattern of the CMB anisotropy power spectrum, effectively acting
as a reduction of the baryon fraction. We find that the scalar
magnetic energy density perturbation contribution to the CMB
temperature anisotropy is small compared to the contribution to
the CMB $E$-polarization anisotropy.
\end{abstract}

\maketitle
\section{Introduction}
A promising explanation for  observed uniform magnetic fields in
galaxies is that they are the amplified remnants of a seed
 cosmological
magnetic field  (for reviews see Refs. \cite{widrow02})  generated
in the early Universe  \cite{generation,ratra92}. A seed magnetic
field may have an helical part \cite{helicity-generation}.
Magnetic helicity plays an important role in
magnetohydrodynamical processes in the primordial plasma as well
as in cosmological perturbation dynamics; In particular, magnetic
helicity influences the inverse cascade mechanism --- when
energy is transferred from small to large scales ---
\cite{inverse}, and as a result affects large-scale magnetic
field formation \cite{mhd}.

The average energy density and  helicity of the magnetic field
must be small  to be consistent with the observed large-scale
spatial isotropy of the Universe. In this case the linear theory
of gravitational instability can be used to study perturbation
dynamics \cite{lifshitz,bardeen80,giovannini04,giovannini05}. A
cosmological magnetic field induces scalar, vector and tensor
perturbations \cite{giovannini04}. At linear order each mode
evolves separately. (At second order the modes are coupled and
this results in non-Gaussian effects \cite{brown05}.)

The vector (vorticity) and tensor (gravitational waves)
perturbation modes
 induced by a cosmological magnetic field have attracted
a lot of interest \cite{vector,tensor,mkk02,
cdk04,lewis04,gang04,kr05,giovannini06,tsb05}. This is partially
because they give rise to a $B$-polarization  CMB anisotropy
signal,  which vanishes for density (scalar mode) perturbations
at linear order.  Any cosmological signature of a primordial
magnetic field is a potential probe (for a short review see
Ref.~\cite{subr06}). For example, the limit on a
chemical-potential-like distortion of the CMB Planck spectrum
leads to a limit on a cosmological magnetic field of order
$10^{-8}-10^{-9}$ Gauss on $1-500$ kpc length-scales
\cite{jedamzik2000}. Similar  limits on a cosmological
 magnetic field generated during inflation \cite{ratra92} are obtained  from
CMB temperature and polarization anisotropy and non-gaussianity
data from vector and tensor perturbation modes \cite{lewis04b}.

On the other hand, the scalar mode of magnetically driven
perturbations also has a significant effect on CMB fluctuations:
fast magnetosonic waves shift the CMB temperature anisotropy
power spectrum acoustic peaks \cite{Adams,grasso,y05}. In this
paper we present a systematic treatment of scalar magnetized
perturbations that complements earlier work
\cite{lemoine,Adams,jedamzik98,BS1,grasso,koh,giovannini05,y05,giovannini06}.
Using the total angular momentum formalism \cite{hu97} and
analytical approximation techniques described in \cite{mkk02} we
obtain analytical expressions for CMB temperature and
polarization  anisotropies.

Reference \cite{koh} presents numerical computations of  scalar
 CMB temperature and polarization anisotropies in the case of a stochastic
 cosmological magnetic field with magnetic
 field power spectral indices
$n_B=1, 2, 3$. Here we consider a general cosmological magnetic
field and contrary
 to Ref.~\cite{koh} we account for the
  Lorentz force term
in the Euler equation for baryons, in accord with the analyses of
Refs.~\cite{giovannini04,giovannini05,giovannini06}. The main new
results are approximate analytical expressions for the CMB
fluctuations. This analysis allows us to identify two different
effects arising from the magnetic field: (i) a rescaling of the
photon-baryon fluid sound speed (that is responsible for the
shift of the CMB acoustic peaks); and, (ii) effects from non-zero
magnetic anisotropic stress (that is responsible for the
additional CMB $E$-polarization anisotropy signal).

The rest of this paper is organized as follows. In the next
Section we describe the magnetic field, including the power
spectrum, the anisotropic stress tensor, and its connection with
the scalar  (longitudial) part of the Lorentz force. We present
analytical expressions for two-point correlations functions of
the magnetic field energy density, the scalar part of the Lorentz
force, and the magnetic field anisotropic stress.   In Sec.~III we
derive the equations that govern  scalar magnetic perturbations
(the Einstein and matter conservation equations) and discuss
general solutions and the initial conditions we adopt. In
Secs.~IV---VI we use semi-analytical approximations to compute CMB
temperature and polarization anisotropies (as well as
cross-correlations between temperature and $E$-polarization
anisotropies). We conclude in Sec.~VII.

\section{Magnetic field statistical properties}
\subsection{Power spectrum}

We assume the presence of a
 Gaussianly-distributed stochastic helical cosmological magnetic
 field  generated during or
prior to the radiation-dominated epoch, with the energy density of the
 field a first-order perturbation to the
Friedmann-Lema\^\i tre-Robertson-Walker (FLRW) homogeneous
cosmological spacetime model. We neglect fluid back-reaction onto
the magnetic field, therefore the spatial and temporal dependence
of the field separates, ${\mathbf B}(t,{\mathbf x})={\mathbf
B}({\mathbf x})/a^2$. Here $a(t)$ is the
 cosmological scale factor, normalized to unity at the present time $t_0$,
and ${\mathbf B}({\mathbf x})$ is the magnetic field at the
present time. Since the magnetic field energy density is first
order, the magnetic field is $1/2$ order.

Smoothing on a comoving
 length $\lambda$ with a
Gaussian smoothing kernel $\propto \mbox{exp}[-x^2/\lambda^2]$,
 we obtain the smoothed
magnetic field with mean  squared magnetic field ${B_\lambda}^2 =
\langle {\mathbf B}({\mathbf x}) \cdot {\mathbf B}({\mathbf
x})\rangle |_\lambda$ and mean squared magnetic helicity
${H_\lambda}^2 = \lambda | \langle{\mathbf B}({\mathbf x}) \cdot
[{\mathbf \nabla} \times {\mathbf B} ({\mathbf x})]
\rangle|_\lambda$ \cite{cdk04,kr05}. Corresponding to the
smoothing length $\lambda$ is the smoothing wavenumber
$k_\lambda=2\pi/\lambda$. We use
\begin{equation}
   B_j({\mathbf k}) = \int d^3\!x \,
   e^{i{\mathbf k}\cdot {\mathbf x}} B_j({\mathbf x}),~~~~~~~~~~~
   B_j({\mathbf x}) = \int {d^3\!k \over (2\pi)^3}
   e^{-i{\mathbf k}\cdot {\mathbf x}} B_j({\mathbf k}),\end{equation}
when Fourier transforming between position and wavenumber spaces.
We assume flat spatial hypersurfaces
(consistent with current observational indications, e.g., Ref.~\cite{pr03}).

We
 also assume that the primordial plasma is a perfect conductor on
all
scales larger than the Silk damping wavelength
$\lambda_S$ (the thickness of the last scattering surface)
set by photon and neutrino diffusion. On much smaller scales
we model magnetic field  damping by an ultraviolet cut-off wavenumber
$k_D=2\pi/\lambda_D$ that is due to the damping of
Alfv\'en waves from photon viscosity \cite{jedamzik98,BS1}
(see Eq.~(1) of Ref.~\cite{kr05}); here $\lambda_D \ll \lambda_S$.

Under these assumptions the magnetic field two-point correlation function
in wavenumber space is
 \begin{equation}
 \langle B^\star_i({\mathbf k})B_j({\mathbf k'})\rangle
=(2\pi)^3 \delta^{(3)}
({\mathbf k}-{\mathbf k'}) [P_{ij}({\mathbf{\hat k}}) P_B(k)  +
i \epsilon_{ijl} \hat{k}_l P_H(k)].
\label{spectrum}
\end{equation}
Here $i$ and $j$ are spatial indices, $i,j \in (1,2,3)$, $\hat{k}_i=k_i/k$  a unit wavevector,
$P_{ij}({\mathbf{\hat k}})=\delta_{ij}-\hat{k}_i\hat{k}_j$
 the transverse plane projector, $\epsilon_{ijl}$  the antisymmetric
symbol, and
 $\delta^{(3)}({\mathbf k}-{\mathbf k'})$  the Dirac delta function.
 The power spectra of the
symmetric and helical parts of the magnetic field, $P_B(k)$ and
$P_H(k)$, are assumed to be simple power laws on large scales,
\begin{equation}
P_B(k) = P_{B0}k^{n_B}= \frac{2\pi^2 \lambda^3
B^2_\lambda}{\Gamma(n_B/2+3/2)} (\lambda k)^{n_B},~~~~~~~~~~
P_H(k) = P_{H0}k^{n_H}= \frac{ 2\pi^2 \lambda^3
H^2_\lambda}{\Gamma(n_H/2+2)} (\lambda k)^{n_H},\qquad k<k_D,
\label{energy-spectrum-H}
\end{equation}
and vanish on small scales where $k>k_D$. The spectral indexes
$n_B$ and $n_H$ are constrained by the requirement of finiteness of mean
 magnetic
field energy density ($n_B>-3$) and mean magnetic helicity ($n_H
>-4$) in the infrared region at small $k$. In addition, causality
requires $P_B(k) \geq |P_H(k)|$ (the Schwartz inequality),
\cite{L}.

\subsection{Anisotropic stress-energy  tensor}
We consider the effects of a cosmological magnetic field on the
CMB at high redshift when the Universe is hot and the plasma
 a good conductor. As a result the magnetic field lines are
 dragged by the matter fluid and this generates a weak ``frozen''
 electric field  ${\mathbf E}=-{\bf v \times B}$, where ${\bf v}$
 is the perturbed (first order) 3-velocity of the fluid.
We neglect this weak electric  field in what follows
 since  the energy density of
this  electric field contributes at third order in the
perturbation expansion. At the current time,
 the space-space part of  Maxwell stress-energy
 tensor for the magnetic
field is
\begin{equation}
\tau^{(B)}_{ij}({\mathbf x}, \eta_0)=\frac{1}{4\pi}\left[ B_i({\mathbf
x})B_j({\mathbf x})-\frac{1}{2}\delta_{ij}B^2({\mathbf x})\right ],
\end{equation}
where $\eta_0$ is the current value of conformal time
$\eta = \int^\eta dt/a(t)$.
The energy density and the  anisotropic
 trace-free part of the space-space components of the stress-energy tensor
of the magnetic field at the current time are
\begin{eqnarray}
\rho_B ({\bf x}, \eta_0) &=& \frac{1}{8\pi} B^2 ({\bf x}) =
-\tau^{\rm{tr}}({\mathbf x}, \eta_0),
\label{r}\\
\tau^{(A)}_{ij}({\mathbf x}, \eta_0)&=&\frac{1}{4\pi}\left[
B_i({\mathbf
x})B_j({\mathbf x})-\frac{1}{3}\delta_{ij}B^2({\mathbf x}) \right ] =
\tau^{(B)}_{ij}({\mathbf x}, \eta_0) - \frac{1}{3}\delta_{ij}
\tau^{\rm{tr}}({\mathbf x}, \eta_0)
,
\label{eq:Bx-stress}
\end{eqnarray}
where $\tau^{\rm{tr}} = \delta_{ij} \tau_{ij}^{(B)}$. Both
 $\rho_B({\bf x}, \eta_0) =\rho_B({\bf x}, \eta) a^4 $ and
$\tau^{(A)}_{ij}({\bf x}, \eta_0) =\tau^{(A)}_{ij}({\bf x}, \eta)a^4 $
 are quadratic in the
 magnetic
field,  and  their wavenumber space transforms are convolutions of
the magnetic field,
\begin{eqnarray}
\rho_B({\bf k}, \eta_0)&=& - \tau^{\rm{tr}}({\mathbf k}, \eta_0)=\frac{1}{8\pi} \int
\frac{d^3\!p}{(2\pi)^3} B_{l}({\bf p})B_l({\bf k-p}),
\label{rk}
\\
\tau^{(A)}_{ij}({\mathbf k}, \eta_0)&=&\frac{1}{4\pi} \int
\frac{d^3\!p}{(2\pi)^3} \left[B_{i}({\bf p})B_j({\bf k-p})-
\frac{1}{3}\delta_{ij}B_l({\bf p})B_l({\bf k-p})\right].
\label{eq:Bk-stress}
\end{eqnarray}

We assume that Eq.~(\ref{eq:Bx-stress})
is a
first order perturbation and
 decompose it into   scalar,
vector, and tensor parts, $\tau^{(A)}_{ij} ({\bf x}) =
\Pi_{ij}^{(S)} ({\bf x})+ \Pi_{ij}^{(V)}({\bf
x})+\Pi_{ij}^{(T)}({\bf x})$ \cite{lifshitz,lemoine}. This
decomposition is more conveniently  done in wavenumber space.
 We use  projection operators and find
$ \Pi_{ij}^{(V)}({\mathbf{k}})=[P_{ib}({\mathbf{\hat k}})
\hat{k}_j+P_{jb}({\mathbf{\hat k}})\hat{k}_i]\hat{k}_a
\tau_{ab}({\mathbf{k}}) $ (for the vector part)
 \cite{mkk02,kr05} and  $
\Pi_{ij}^{(T)} ({\mathbf {k}}, \eta_0)=[ P_{ia}({\mathbf{\hat
k}})P_{jb}({\mathbf{\hat k}}) - P_{ij}({\mathbf{\hat
k}}) P_{ab}({\mathbf{\hat k}})/2 ]\tau_{ab}({\mathbf{ k}}, \eta_0)$
(for the tensor part) \cite{mkk02,cdk04}. The scalar parts of
 $\rho_B$ and $\tau_{ij}^{(B)} $
determine the scalar part of the magnetic source.
 The scalar part $\Pi^{(S)}_{ij}({\bf k})$ has to be
 proportional to ${\hat k}_i
{\hat k}_j - \delta_{ij}/3$
\cite{lifshitz}, so  we define
\begin{equation}
 \Pi^{(S)}_{ij}({\bf k})= \frac{3}{2}\left({\hat k}_i {\hat k}_j -
\frac{1}{3}\delta_{ij} \right) \Pi^{(S)}({\bf k}).
\end{equation} Here
the scalar $\Pi^{(S)}({\bf k}, \eta)=\Pi^{(S)}({\bf k},
\eta_0)/a^4$ is associated with the anisotropic stress of the
magnetic field. (At  leading order the isotropic pressure has the
same time dependence, $p_B({\bf x}, \eta) = p_B({\bf x},
\eta_0)/a^4$,  and is related to the magnetic field energy density
 by  $p_B({\bf x}, \eta) = \rho_B({\bf x}, \eta)/3$.)
 It is straightforward to determine $\Pi^{(S)}({\bf k})$ by applying
${\hat k}_n {\hat k}_m - \delta_{nm}/3$ on $\tau_{nm}^{(A)}({\bf
k})$, i.e.,
\begin{equation}
\Pi^{(S)}({\bf k})
= {\hat k}_n {\hat k}_m\tau_{nm}^{(A)}({\bf k}),
\label{pi22}
\end{equation}
where we use $\delta_{nm} \tau^{(A)}_{nm}=0$. Our
 $\Pi^{(S)}({\bf k}, \eta_0)$ is related to the $\tau^{(S)}({\bf k}, \eta_0)$
of Ref.~\cite{brown05} through
$\tau^{(S)}({\bf k}, \eta_0)=3\Pi^{(S)}({\bf k}, \eta_0)/2$.

The scalar $\Pi^{(S)}$ is related to  the scalar part of the
Lorentz force ${\bf L} ({\bf x}, \eta_0)=-[{\bf B}({\bf x}) \times
({\bf \nabla} \times {\bf B}({\bf x}))]/(4\pi)$ and the isotropic
pressure $p_B({\bf x}, \eta_0)$. We introduce a  scalar
$L^{(S)}$ defined by  $ L_i^{(S)}({\bf x}, \eta_0) =
\nabla_i L^{(S)} ({\bf x}, \eta_0)$, where $L_i^{(S)}$ is the scalar part of the Lorentz force. Using the Maxwell equation
${\mathbf \nabla} \cdot {\mathbf B}=0$, the Lorentz force  is
\begin{equation}
L_i ({\bf x}, \eta_0) = \frac{1}{4\pi} \left[ B_j({\bf x})
\nabla_j B_i ({\bf x}) - \frac{1}{2} \nabla_i B^2 ({\bf x})
\right], \label{li}
\end{equation}
and the corresponding scalar part is derived through the scalar
\begin{equation}
\nabla^2 L^{(S)} ({\bf x}, \eta_0) = \nabla_i L^{(S)}_i({\bf x}, \eta_0) =
 \frac{1}{4\pi} \left[
(\nabla_i B_j({\bf x})) \nabla_j B_i ({\bf x})- \frac{1}{2}
\nabla^2 B^2 ({\bf x})
\right],
\label{12}
\end{equation}
where $\nabla^2 = \nabla_i \nabla _i$ is the Laplace operator and we have used
${\mathbf \nabla} \cdot {\mathbf B}=0$.
 In position space Eq.~(\ref{pi22}) reads
\begin{equation}
\nabla^2\Pi^{(S)} ({\bf x}, \eta_0) = {1 \over 4 \pi}
\left[\nabla_i \nabla_j (B_i({\bf x}) B_j ({\bf x})) - {1 \over
3}\nabla^2  B^2({\bf x}) \right],
\end{equation}
and comparison with Eq.~(\ref{12}) results in
\begin{equation}
\Pi^{(S)} ({\bf x},\eta_0) =
\frac{\rho_B({\bf x},\eta_0)}{3} + L^{(S)}({\bf x}, \eta_0),
\label{relation}
\end{equation}
the analog of Eq.~(5.44) of Ref.~\cite{giovannini05}.

Since we consider a stochastic magnetic field we present here
various correlations and averages of the magnetic source for
scalar perturbations. The wavenumber-space scalar two-point
correlation function is \begin{eqnarray}
\langle\Pi^{(S)\star}({\mathbf k}, \eta_0) \Pi^{(S)}({\mathbf
k^\prime}, \eta_0) \rangle=(2\pi)^3|\Pi^{(S)}({k}, \eta_0)|^2
\delta^{(3)} ({\mathbf k} - {\mathbf k^\prime}), \end{eqnarray}
where the power spectrum $|\Pi^{(S)}({k}, \eta_0)|^2$ depends
only on
 $k=|{\bf k}|$. (The vector and tensor two-point correlation functions are given
in Refs.~\cite{mkk02,cdk04,lewis04,kr05}.)
Two-point correlation functions of the
magnetic field energy density and
 the scalar part of the Lorentz force are defined in a similar manner,
\begin{eqnarray}
\langle\rho_B^{\star}({\mathbf k}, \eta_0) \rho_B({\mathbf
k^\prime}, \eta_0) \rangle &= & (2\pi)^3|\rho_B({k}, \eta_0)|^2
\delta^{(3)} ({\mathbf k} - {\mathbf k^\prime}),
\\
\langle L^{(S)\star}({\mathbf k}, \eta_0) L^{(S)}({\mathbf
k^\prime}, \eta_0) \rangle &= & (2\pi)^3|L^{(S)}({k}, \eta_0)|^2
\delta^{(3)} ({\mathbf k} - {\mathbf k^\prime}). \end{eqnarray}
Note that  $L^{(S)}({\bf x}, \eta)$ is related to ${\mathcal
F}({\bf x}, \eta)$ of Ref.~\cite{giovannini04} through $\nabla^2
L^{(S)}({\bf x}, \eta)={\mathcal F}({\bf x}, \eta)/(4\pi)$, so
$k^2L^{(S)}({\bf k}, \eta)=-{\mathcal F}({\bf k}, \eta)/(4\pi)$.

The scalar power spectra $|\Pi^{(S)}(k)|^2$, $|\rho_B(k)|^2$,  and
$|L^{(S)}(k)|^2$ are determined by   the symmetric  part of the
magnetic field power spectrum $P_B(k)$ and do not depend on the
magnetic helicity spectrum $P_H(k)$ (also see
Ref.~\cite{brown05}),
\begin{eqnarray}
|\Pi^{(S)}(k)|^2 &=&  \frac{1}{576\pi^5}
\int d^3\!p~P_B(p) P_B(|{\bf k-p}|) \left[9(1-\gamma^2)(1-\beta^2) -
6(1+\gamma\mu\beta-\gamma^2-\beta^2) + (1+\mu^2) \right],
\label{pis-spectrum}\\
|\rho_B(k)|^2 &=& \frac{1}{256\pi^5}
\int d^3\!p~P_B(p) P_B(|{\bf k-p}|) (1+\mu^2),
\label{rho-spectrum}\\
|L^{(S)}(k)|^2 &=&  \frac{1}{256\pi^5}
\int d^3\!p~P_B(p)
P_B(|{\bf k-p}|) \left[4\gamma\beta (\gamma\beta -\mu) + (1+\mu^2)\right].
\end{eqnarray}
Here $\gamma = {\hat{\bf k}}\cdot {\hat{\bf p}}$,
$\beta = {\bf k}\cdot ({\bf{k-p}})/(k|{\bf{k-p}}|)$, and
$\mu = {\bf p}\cdot ({\bf{k-p}})/(p|{\bf{k-p}}|)$.
The relations to the two-point correlation functions given in Ref.
\cite{brown05} are  $\langle \tau^\star ({\bf k}) \tau
({\bf k'}) \rangle = \langle\rho_B^{\star}({\mathbf k}) \rho_B({\mathbf
k^\prime}) \rangle$  and $\langle \tau^{\star S} ({\bf k}) \tau^{S}
({\bf k'})\rangle = 9\langle\Pi^{(S)\star}({\mathbf k}) \Pi^{(S)}({\mathbf
k^\prime}) \rangle/4$; $\tau({\bf k})$ and
$\tau^{S} ({\bf k})$ are given in Eqs.~(2.15) of Ref.~\cite{brown05}.

Using the power law magnetic field power spectrum of
Eq.~(\ref{energy-spectrum-H}), we can obtain expressions for the
power spectra $|\Pi^{(S)}(k)|^2$, $|\rho_B(k)|^2$, and
$|L^{(S)}(k)|^2$ in the  semi-analytical approximation where we
divide the integration range into $p < k$ and $p>k$ parts and
consider  two limiting ranges $p \ll k$ and $p \gg k$
\cite{mkk02,cdk04,kr05}. In particular, the magnetic energy
density correlation power spectrum is{\footnote{The power spectrum
$|\rho_B (k)|^2$ is related to the two-point correlation function
of the energy density of the magnetic field.
 By  definition the r.m.s.\ magnetic field
energy density ${\rho_B}^{\rm{rms}} $ is
\begin{equation}
[{\rho_B}^{\rm{rms}}(\eta_0)]^2 = \langle
\rho_B({\bf x}) \rho_B({\bf x}) \rangle = \frac{1}{2\pi^2}
\int_0^\infty dk~k^2 |\rho_B (k, \eta_0)|^2.
\end{equation}
Replacing the upper limit of integration by the cut-off scale
$k_D$
 and using Eq.~(\ref{pi-sca}) below, we find
\begin{equation}
  {\rho_B}^{\rm{rms}}(\eta_0)
= \frac{\sqrt{n+6}
B_\lambda^2 (k_D \lambda)^{n_B+3}}{8\sqrt{2}(n+3)\pi \Gamma(n_B/2+3/2)}.
\end{equation}
${\rho_B}^{\rm{rms}}(\eta_0)$ differs from the average magnetic energy
density ${\bar\rho}_B = \langle \rho_B \rangle $
which is determined by the power spectrum $E_B(k)=k^2P_B(k)/\pi^2$,
\begin{equation}
{\bar\rho_B}(\eta_0) = \frac{1}{8\pi} \langle B_l({\bf x}) B_l({\bf x})
\rangle = \frac{1}{8\pi}
\int_0^\infty dk E_B(k),
\end{equation}
which, using the cut-off scale $k_D$, gives
\begin{equation}
  {\bar\rho}_B(\eta_0)
= \frac{B_\lambda^2 (k_D \lambda)^{n_B+3}}{4\pi(n_B+3)
\Gamma(n_B/2+3/2)};
\end{equation}
for $n_B = -3$, $ {\bar\rho}_B(\eta_0) = {B_\lambda^2}/(8\pi)$. We
note, in particular, that the average Lorentz force vanishes but
the r.m.s.\ Lorentz force is not zero. }}
\begin{eqnarray}
|\rho_B({k}, \eta_0)|^2 &=& \frac{3 (k_D\lambda)^{2n_B+3}
 \lambda^3 B_\lambda^4}{32  (2n_B+3) \Gamma^2(n_B/2+3/2)}
\left[
        1 +\frac{n_B}{n_B+3}\left(\frac{k}{k_D}\right)^{2n_B+3}\right].
\label{pi-sca} \end{eqnarray}
For $n_B>-3/2$ this expression  is
 dominated by the cut-off
scale $k_D$, and for large $k_D$ it does not depend on $k$,
 while for $n_B<3/2$ we get
$|\rho_B(k, \eta_0)|^2 \propto k^{2n_B+3}$.  It can be shown
\cite{mkk02} that $|\rho_B({k}, \eta_0)|^2 \simeq 9 |L^{(S)}({k},
\eta_0)|^2/8 \simeq |L^{(S)}({k}, \eta_0)|^2$  and $|\rho_B({k},
\eta_0)|^2 \simeq 9 |\Pi^{(S)}({k}, \eta_0)|^2/4$.{\footnote{ It
may be shown  that $L^{(S)}\simeq -2\sqrt{2}\rho_B/3\simeq -\rho_B
$ and $\Pi^{(S)} \simeq - 2\rho_B/3$. }}

Increasing magnetic helicity reduces the vector part of Lorentz
force two-point correlation function $\langle L^{((V)}_i L^{(V)}_i
\rangle$ (\cite{giovannini04,kr05};  this reduces parity-even CMB
fluctuations \cite{cdk04,kr05}), but leaves the scalar part
unchanged ($|L^{(S)}|^2$ is independent of magnetic helicity). In
contrast, Refs.~\cite{koh,giovannini04,giovannini05} neglect the
Lorentz force for a maximally helical magnetic field. They argue
as follows: since $\langle {\bf B} \cdot ({\bf \nabla \times B})
\rangle$ is  maximal, the average of the Lorentz force $\propto
\langle {\bf B} \times ({\bf \nabla \times B}) \rangle$ for a such
field is minimal or even zero --- this is a valid approximation
for a homogeneous field, and  results in the force-free
approximation \cite{giovannini04} --- but this  is not applicable
for a stochastic field. In the case of a stochastic field  the
average Lorentz force is zero (as is  the average magnetic field
itself) but  the  Lorentz force two-point correlation is non-zero
(see footnote 2 above). This affects stochastic peculiar motions
(vorticity perturbations) of
 charged particles \cite{kr05}, and,  as we show below, the  dynamics of
 density perturbations also. In the
stochastic field case the force-free approximation should be used with caution.

\section{Scalar magnetic perturbations}

In this section we study the dynamics of  linear magnetic energy
density perturbations about a spatially-flat FLRW background with
scalar metric fluctuations. The metric tensor can be decomposed
into a spatially homogeneous background part and a perturbation
part, $g_{\mu\nu}=g^{(0)}_{\mu\nu}+\delta g_{\mu\nu}$; Greek
letters are used for spacetime indices, $\mu ,\nu \in (0, 1, 2,
3)$. For a spatially-flat model, and working with conformal time,
the background  FLRW metric $g^{(0)}_{\mu\nu}=a^2\eta_{\mu\nu}$,
where $\eta_{\mu\nu}=\mbox{diag}(-1,1,1,1)$ is the Minkowski
metric tensor. Scalar perturbations are gauge dependent because
the mapping of coordinates between the perturbed physical
manifold and the background is not unique. We work  in Newtonian
(longitudinal) gauge in which the metric tensor is shear free
\cite{bardeen80}. Scalar perturbations to the geometry are then
described by two scalar gravitational potentials $\Psi$ and
$\Phi$ where
\begin{equation}
\delta g^{(S)}_{00}=-2a^2\Psi,\qquad\qquad
\delta g^{(S)}_{ij}=2a^2\Phi\delta_{ij}.
\label{eq:S-metric-pert}
\end{equation}
In this gauge, neglecting vector and tensor fluctuations,  the line element $ds^2 = a^2(\eta)[-(1+2\Psi) d\eta^2
+ \delta_{ij} (1+ 2\Phi) dx^i dx^j]$.

Matter perturbations are described by the perturbation
of the complete  stress-energy tensor, $\delta \tau_\mu^\nu$, which is the
 sum of the perturbed fluid and electromagnetic stress-energy tensors,
$\delta \tau_\mu^\nu = \sum_f \delta \tau_{{f} \mu}^\nu +
\tau_{\mu}^{(B)\nu}$. The subscript $f$ denotes the three
different
 matter components we consider, photons ($\gamma$), baryons ($b$), or cold dark matter ($c$),
 which we model as fluids. A  magnetic field influences perturbations
without changing the background metric which is determined
by the photon, baryon, and cold dark matter background densities.
 For simplicity we ignore neutrinos; Refs.~\cite{lewis04b,giovannini04,giovannini05} account for
the effects of  these relativistic weakly interacting
particles. Since we focus on dynamics at  large  redshift we also ignore
 a possible cosmological constant or  dark energy. We decompose
 perturbations into plane waves
$\propto {\rm exp}({i~{\mathbf k} \cdot {\mathbf x}})$ and in
what follows equations are presented in wavenumber space.

The magnetic field source affects the motion of baryons.
Before recombination the photon-baryon plasma
is dominated by photons and has the relativistic equation of state
 $p=\rho/3$. The baryon-photon momentum density ratio
$R(\eta)=3\rho_b(\eta)/(4\rho_\gamma(\eta))=
3a \Omega_{0b}/(4\Omega_{0\gamma})$,
where $\rho_\gamma$ and $\rho_b$ are the energy densities of
photons and  baryons,
 $\Omega_{0\gamma}$ and $\Omega_{0b}$ are the photon and baryon
density parameters measured today,
i.e., $\Omega_{0f} = \rho_f(\eta_0)/\rho_{\rm{cr}}$,
where $\rho_{\rm{cr}} = 8\pi G/(3H_0^2)$ is the critical Einstein-de Sitter
 density  and $H_0$ is the Hubble constant. At early times $R\ll 1$;
at the last scattering surface when photons and baryons decouple
$R_{\rm{dec}} \simeq 0.35$.

We consider a primordial magnetic field  and  assume that the
stress-energy of the magnetic field is not compensated by
anisotropic stress in the fluid. That is,  we have non-zero
initial  gravitational potentials but vanishing initial
 fluid energy density  and velocity perturbations.
This is possible for a magnetic field generated during inflation
\cite{ratra92,lemoine}. This will limit the spectral index of the
magnetic field, $n_B < 0$.\footnote{ This limit on the magnetic
field spectral index was  obtained
 for a  magnetic field generated through coupling of the inflation and
 hypercharge during inflation \cite{ratra92}. Ref.~\cite{hogan} argues
for the same result while Ref.~\cite{dc} relaxes the limit to  $n_B<2$.}

In  subsections A and B below we present the linear perturbation
theory
equations for the metric and matter perturbations. 
In subsection C we discuss the speed of sound rescaling in the
presence of a magnetic field, while in subsection D we consider
initial conditions.

\subsection{Metric perturbations}
In zero-shear (Newtonian) gauge,  part of Einstein's equation for
scalar metric perturbations become
 two Poisson equations \cite{lifshitz,bardeen80}, which in wavenumber space
 are
\begin{eqnarray}
&&k^2 \Phi
= 4 \pi G a^2 \left[\rho_B   +\sum_{f}\rho_f \delta_f  +
3 \frac{{\dot a}}{a k} \sum_f (\rho_f + p_f) v^{(S)}_f \right ],
\label{phi} \\
&&k^2(\Psi + \Phi) = - 4 \pi G a^2 \left[ 2 \sum_f p_f \Pi_f^{(S)}
+ \rho_B + 3L^{(S)} \right].  \label{psi+phi}
\end{eqnarray}
Here  $\delta_f$ is the $f$-th fluid density perturbation,
 $v_f^{(S)}$ is the scalar part of the $f$-th fluid velocity
perturbation ${\mathbf v_f^{(S)}}={\mathbf {\hat k}} v^{(S)}_f$,
 $\Pi^{(S)}_f$ is the scalar part of the anisotropic $f$-th fluid
 stress-energy tensor,
$\Pi^{(S)}_f=({\hat k}_i {\hat k}_j - \delta_{ij}/3) \tau_{f,
ij}$, and an overdot represents a conformal time derivative
$\partial/\partial \eta$.
 For economy of notation we do not explicitly show the
wavevector (${\bf k}$) or time ($\eta$) dependence of the
variables in these and following equations. (For instance,
 the magnetic field energy density $\rho_B (\eta)=
\rho_B(\eta_0)/a^4$, the Lorentz force $L^{(S)} (\eta) =
L^{(S)}(\eta_0)/a^4$, and  the anisotropic stress of the magnetic
field $\Pi^{(S)} (\eta) = \Pi^{(S)}(\eta_0)/a^4$.)
 Note that the
combination on the r.h.s.\ of Eq.~(\ref{phi}),
\begin{equation}
{\mathcal D}_f = \rho_f \delta_f  + 3 \frac{{\dot a}}{a k} (\rho_f
+ p_f) v^{(S)}_f  = \rho_f \left[ \delta_f + 3 \frac{{\dot a}}{a}
(1+\omega_f) \frac{v_f^{(S)}}{k}\right], \label{DD}
\end{equation}
(where  $\omega_f = p_f/\rho_f$ is the equation of state
parameter for the $f$-th fluid) corresponds to the gauge-invariant
 total (magnetic-field-induced) $f$-th
fluid energy density perturbation \cite{kodama}.

To derive Eq.~(\ref{phi}) we use the Einstein equation for
$\delta\tau^{i}_0$, which is
\begin{equation}
- {\dot \Phi} + \frac{{\dot a}}{a}\Psi  = 4\pi G a^2
\sum_f (\rho_f +p_f) v_f^{(S)}/k.
\label{1}
\end{equation}
As a consequence of the high conductivity, $\sigma \gg 1 $, of the
primordial plasma,  the $0i$ component of the magnetic field
stress-energy tensor, $\propto {\bf E} \cdot {\bf B}$, being
suppressed by $1/\sigma$, does not contribute to the  r.h.s.\ of
Eq.~(\ref{1}). The energy density of the electric field, $\propto
E^2$, is suppressed by $1/\sigma^2$  and
 does not contribute to the  r.h.s.\ of Eq.~(\ref{phi}).

The trace of the space-space part of the Einstein equation gives an
 additional constraint equation. Since this  equation
 is for the trace $\tau^i_i$ \cite{koh,giovannini04},
 it does not have a   contribution from
magnetic anisotropic stress (but isotropic
pressure does contribute),
\begin{equation}
{\ddot \Phi} + \frac{{\dot a}}{a} \left(2{\dot \Phi} - {\dot
\Psi}\right) + \left[ \left(\frac{{\dot a}}{a}\right)^2 -
2\frac{{\ddot a}}{a} \right]\Psi +\frac{k^2}{3}\left(\Psi + \Phi
\right) = -4\pi Ga^2 \left( \sum_f c_{S,f}^2 \rho_f \delta_f
+\frac{\rho_B}{3} \right).
 \label{2}
\end{equation}
Here $c_{S, f}^2 = dp_f/d\rho_f$ is the square of the speed of
sound in the $f$-th fluid.

 Equations (\ref{phi}), (\ref{psi+phi}),  and (\ref{2})
 govern the evolution of the scalar
metric perturbations $\Phi$ and $\Psi$,  if the density,
velocity, and anisotropic stress perturbations for each $f$-th
component are known.
 For unmagnetized perturbations, the energy density and anisotropic stress
of the
magnetic field, $\rho_B({\bf k}, \eta)$ and  $\Pi^{(S)}({\bf k}, \eta)$,
 vanish.
In this case Eq. ({\ref{psi+phi}) results in  the  simple relation
$\Phi=-\Psi$. The presence of collisionless particles, such as
neutrinos, induces anisotropic stress \cite{hu97,lewis04b}.
 Even through we neglect neutrinos, in our case
a stochastic magnetic field induces anisotropic stress and so  violates
the condition $\Phi=-\Psi$.

Using Eq.~(\ref{psi+phi}), setting $p_b=0=p_c$,
and neglecting neutrinos, Eq.~(\ref{2}) becomes
\begin{equation}
{\ddot \Phi} + \frac{{\dot a}}{a} \left(2{\dot \Phi} - {\dot \Psi}\right)
+ \left[ \left(\frac{{\dot a}}{a}\right)^2 -
2\frac{{\ddot a}}{a} \right]\Psi = -\frac{4\pi Ga^2\rho_\gamma}{3}
\left(\delta_\gamma - \frac{3L^{(S)}}{\rho_\gamma} \right)=
-\frac{{\dot a}^2}{2a^2}
\left(\delta_\gamma - \frac{3L^{(S)}}{\rho_\gamma} \right),
\label{22}
\end{equation}
where the last step uses the zeroth order Friedmann equation in the radiation dominated model.
Equations (\ref{phi}) and (\ref{psi+phi}) may be combined together,
\begin{equation}
k^2 (2\Phi +\Psi)
= 4 \pi G a^2 \left[ \rho_\gamma \left(
\delta_\gamma - \frac{3 L^{(S)}}{\rho_\gamma}\right)
+\rho_b \delta_b+\rho_c \delta_c  +
3 \frac{{\dot a}}{a k} \sum_f (\rho_f + p_f) v^{(S)}_f \right ].
\label{23}
\end{equation}
We show below that the combination  $\delta_\gamma \rho_\gamma -
3L^{(S)}$ on the r.h.s.\ of this equation
 also appears  in the equations for  matter perturbations
  and reflects an  effective temperature rescaling
 (induced by the presence of the Lorentz force).

Using Eq.~(\ref{relation}), Eqs.~(\ref{phi}) and (\ref{psi+phi})
may be rewritten as \begin{eqnarray} &&k^2 \Phi = 4 \pi G a^2
\left[3\Pi^{(S)} + \sum_{f} \rho_f \delta_f  - 3L^{S} + 3
\frac{{\dot a}}{a k} \sum_f (\rho_f + p_f) v^{(S)}_f \right ],
\label{phi1} \\
&&k^2(\Psi + \Phi) = - 12 \pi G a^2 \left[ \frac{2}{3}\sum_f p_f
\Pi_f^{(S)} + \Pi^{(S)} \right]. \label{psi1+phi1}
\end{eqnarray}
These equations  show that we should take note of  two effects,
one a consequence of anisotropic stress, the other
 arising from  non-zero energy density and peculiar velocity
 perturbations.

\subsection{Matter perturbations}

The first-order energy conservation equations for photons,
baryons, and the CDM fluid  are 
\begin{eqnarray}
&&\dot \delta_\gamma +
{4 \over 3} k v^{(S)}_\gamma +4  {\dot \Phi}=0,
\label{density_photon}
\\
&&\dot \delta_b + k v^{(S)}_b +
3 {\dot \Phi}=0,
\label{density_baryons}
\\
&&\dot \delta_{c}  +k v^{(S)}_{c} + 3 {\dot \Phi}=0.
\label{density_cdm}
\end{eqnarray}

The Lorentz force  directly modifies
only the Euler equation for the baryons, since only baryons are charged.
 The Euler equation  for the CDM fluid is identical
to that in a model without a magnetic field. Prior to  decoupling,
photons are tightly coupled to baryons and they move together.
Since the Lorentz force  affects the motion of baryons, the
presence of a cosmological magnetic field also influences the
evolution of photons. The first-order Euler or momentum
conservation equations in the
 tight
coupling regime (prior to  last scattering)
  for photons, baryons, and CDM  are, \cite{Adams,giovannini04},
\begin{eqnarray}
&&{\dot v}_\gamma^{(S)} - { k \over 4} \delta_\gamma  - k \Psi
+ {\dot \tau}
(v^{(S)}_\gamma -v^{(S)}_b) + \frac{2k}{5} \Theta^{(S)}_2=0,
\label{velocity-photon}
\\
&&{\dot v}_b^{(S)} + {\dot
a \over a} v_b^{(S)} - k \Psi -  {\dot \tau \over R} (v^{(S)}_\gamma
-v^{(S)}_b) + {k L^{(S)} \over \rho_b}=0,
\label{velocity-baryon}
\\
&&{\dot v}_{c}^{(S)} + {\dot
a \over a} v_{c}^{(S)} - k \Psi =0.
\label{velocity-cdm}
\end{eqnarray}
In Eq.\ (\ref{velocity-photon}) $\Theta^{(S)}_2 =
5\Pi^{(S)}_\gamma /12$
is  the quadrupole moment of the photon temperature fluctuation
and reflects the anisotropic nature of Thomson-Compton scattering.
 The photon density fluctuation
 is related to the temperature
monopole moment, $\delta_\gamma=4 \Theta_0^{(S)}$, and the
perturbed photon velocity   is the dipole term $v^{(S)}_\gamma=
\Theta_1^{(S)}$. In Eqs. (\ref{velocity-photon}) and
  (\ref{velocity-baryon})
 $\dot \tau$ is the differential
visibility function; $\dot \tau = n_e x_e \sigma_T a$
 where
$n_e(z) $ is the charged particle number density, $x_e(z)$ is the
plasma ionization fraction, and $\sigma_T$ is the Thompson
cross-section. In Eq. (\ref{velocity-baryon}) the term $\propto
{\dot \tau}$ (the so called ``baryon drag force'' term) reflects
the
 coupling
between photons and baryons, and so
 determines the velocity difference between the photon and baryon fluids.
In the lowest order of the tight coupling approximation the $
{\dot \tau} (v^{(S)}_\gamma -v^{(S)}_b)$ terms in Eqs.
(\ref{velocity-photon}) and
 (\ref{velocity-baryon})
 vanish since
$v^{(S)}_b \approx v_\gamma^{(S)}$.

At early time the
hydrodynamical description for photons is a reasonable approximation
due to their strong interaction
 with baryons and short mean free path. In this case  $\Pi_\gamma^{(S)}=0$ and so $\Theta_2^{(S)}=0$.
Subtracting
 Eq.~(\ref{velocity-baryon}) from
Eq.~(\ref{velocity-photon}), multiplying by $R$,
 and using Eq.~(\ref{density_baryons}),
the velocity difference between the  photon and baryon fluids,
 $\Delta v_{\gamma
b} = v^{(S)}_\gamma - v_b^{(S)}$, obeys
\begin{equation}
R \Delta {\dot v}_{\gamma b} + (1+R) {\dot \tau} \Delta v_{\gamma b}
= \frac{k}{4} \left( \frac{3L^{(S)}}{\rho_\gamma} +
R \delta_\gamma \right) - \frac{R{\dot a}}{ka}
({\dot \delta}_b + 3 {\dot \Phi}),
\label{deltav}
\end{equation}
where $\Delta {\dot v}_{\gamma b}= {\dot v}_\gamma^{(S)} -
{\dot v}_b^{(S)}$.   $R \propto a $, so at early times it is very small,
and at these times
 Eq.~(\ref{deltav})
results in{\footnote{For a model without a magnetic field $\Delta
v_{\gamma b}$ obeys
\begin{equation}
R\Delta {\dot v}_{\gamma b} + (1+{R}) {\dot \tau} \Delta v_{\gamma b}
= R \left[\frac {k \delta_\gamma}{4} - \frac{{\dot a}}{ka}
({\dot \delta}_b + 3 {\dot \Phi}) \right],
\label{deltav11}
\end{equation}
and so for $R \ll 1$  $\Delta v_{\gamma
b}$ vanishes.  In the presence of a magnetic field  the
Lorentz force term is responsible for a non-zero velocity difference at early times.}}
\begin{equation}
\Delta v_{\gamma b} \simeq
\frac{3k L^{(S)}}{4{\dot \tau} \rho_\gamma  }.\label{deltav1}
\end{equation}
For modes with wavelengths larger than the Hubble radius, the
difference between photon and baryon fluid velocities is
significantly suppressed by the $k/{\dot \tau}$ factor in this
equation.

To derive an equation that describes the  photon-baryon fluid in the
 tight coupling
approximation, where $v_b^{(S)} \simeq v_\gamma^{(S)} \equiv  v^{(S)}$,
 we multiply
Eq.~(\ref{velocity-baryon}) by $R$ and add
Eq.~(\ref{velocity-photon}) to get
\begin{equation}
\frac{\partial}{\partial \eta}\left[(1+R) v^{(S)}\right] =
\frac{k}{4} \left(\delta_\gamma - \frac{3L^{(S)}}{\rho_\gamma}
\right) + (1+R)k\Psi. \label{vdot}
\end{equation}
Here we have made use of the relation ${\dot a}/a = {\dot R}/{R}$.
Equation (\ref{vdot}) is valid over a limited range of times
prior to decoupling.
 After decoupling photon evolution must be described by the
 Boltzmann transport
equation  and  ${\dot \tau}$ is not large enough
 to ensure the equality of the photon and baryon fluid velocities, i.e.,
$v_\gamma^{(S)} \neq v_b^{(S)}$, so  Eq.~(\ref{vdot}) is not valid.
Also, after decoupling it is no longer possible to ignore the
 photon  anisotropic stress term
(i.e.,  the temperature quadrupole moment $\Pi^{(S)}_\gamma = 12
\Theta^{(S)}_2/5$) that appears in  Eq.~(\ref{velocity-photon}).

Using
 $\delta_\gamma = 4 { \Theta}_0^{(S)}$ and
$v_\gamma^{(S)} = \Theta_1^{(S)}$, Eqs.~ (\ref{density_photon})
and (\ref{vdot})
 may be expressed in terms of temperature multipoles,
\begin{eqnarray}
{\dot \Theta}_0^{(S)} &=&-\frac{k}{3} \Theta_1^{(S)} -  {\dot \Phi},
\label{theta0}
\\
\frac{\partial}{\partial \eta}\left[ (1+R) \Theta_1^{(S)} \right]  &=& k
\left[
\Theta_0^{(S)} - \frac{3L^{(S)}}{4\rho_\gamma}+ (1+R) \Psi \right].
\label{theta1}
\end{eqnarray}
The quantity $\Theta_0^{(S)} + \Psi $ is usually called the
effective temperature}, and in a model with a cosmological
magnetic field it obeys the second order differential equation
\begin{equation}
\frac{\partial}{\partial \eta }\left[
(1+R) ({\dot \Theta_0^{(S)} } + {\dot \Psi} ) \right] +
\frac{k^2}{3} (\Theta_0^{(S)} +\Psi) = - \frac{Rk^2}{3} \Psi +
\frac{\partial}{\partial \eta }
\left[(1+R)({\dot \Psi} - {\dot \Phi}) \right] + \frac{k^2 L^{(S)}}{4
\rho_\gamma}.
\label{thetafin}
\end{equation}
This equation is derived from Eqs.~(\ref{theta0}) and (\ref{theta1}).
In addition to the  usual baryon drag force term
 $\propto Rk^2 \Psi/3$ and
gravitational potential time derivative difference term $~
\propto ({\dot \Psi} - {\dot \Phi})$, there is a  term on the
 r.h.s.\ of this equation, $\propto {k^2 L^{(S)}}/{
\rho_\gamma}$,  which  directly
 reflects the
 presence of
the cosmological magnetic field.

Since $L^{(S)}(\eta)/\rho_\gamma(\eta)$ is time independent, we
may use $\Theta_0^{(S)} + \Psi - 3L^{(S)}/(4\rho_\gamma)$ as the
 generalized effective temperature in the case when a cosmological
 magnetic field is present. Equation (\ref{thetafin}), rewritten in
 terms of ${\bar \Theta}_0^{(S)}$, where $4{\bar \Theta}_0^{(S)}= 4\Theta_0^{(S)} -
3L^{(S)}/\rho_\gamma = \delta_\gamma - 3L^{(S)}/\rho_\gamma$, is
\begin{equation}
\frac{\partial}{\partial \eta} \left[(1+R) ({\dot {\bar
\Theta}_0^{(S) } } + {\dot \Psi} ) \right] + \frac{k^2}{3} ({\bar
\Theta}_0^{(S)} +\Psi) = - \frac{Rk^2}{3} \Psi +
\frac{\partial}{\partial \eta} \left[(1+R)({\dot \Psi} - {\dot
\Phi}) \right]. \label{thetafin1}
\end{equation}
${\bar \Theta}_0^{(S)}$ reflects the rescaling of the photon fluid
energy density perturbation in the presence of a magnetic field.

In a model without a cosmological magnetic field,
 defining $m_{\rm{eff}} = 1+R$,
Eq. (\ref{thetafin}) can be rewritten as (also see Eq. (83) of
Ref. ~\cite{hu97}),
\begin{equation}
\frac{\partial}{\partial \eta }\left( m_{\rm{eff}} {\dot
\Theta_0^{(S)} } \right) + \frac{k^2}{3} \Theta_0^{(S)}  = -
\frac{k^2}{3} m_{\rm{eff}} \Psi - \frac{\partial}{\partial \eta }
\left(m_{\rm{eff}} {\dot \Phi} \right). \label{thetafin111}
\end{equation}
Equation (\ref{thetafin111}) is the second order differential
equation that governs the dynamics of  photon density
perturbations ($\delta_\gamma = 4\Theta_0^{(S)}$). In the absence
of gravitational potentials  the r.h.s.\ of Eq.
(\ref{thetafin111}) vanishes. The l.h.s.\ of
  Eq. (\ref{thetafin111}) differs from the equation for an undriven  simple
 harmonic oscillator only by the time-dependent factor $m_{\rm{eff}}$. Just like the case for a harmonic
 oscillator,
 Eq. (\ref{thetafin111})
 has two independent solutions --- sine and cosine modes --- that
 depend on initial conditions. Defining the
photon-baryon fluid sound speed $c_S = 1/\sqrt {3m_{\rm eff}}$,
for ${\dot m}_{\rm eff}/m_{\rm eff} \ll \omega $  where
$\omega=c_S k$ is the oscillation frequency, the JWKB solutions of
Eqs. (\ref{theta0}) and (\ref{theta1}) are, \cite{hu97},
\begin{eqnarray}
{\Theta}_0^{(S)}   &=& A_1 m_{\rm eff}^{-1/4} \cos(ks +\phi),
\label{theta00}\\
{ \Theta}_1^{(S)}   &=& A_1 \sqrt{3} ~m_{\rm eff}^{-3/4} \sin(ks +
\phi).  \label{theta11}
\end{eqnarray}
Here $A_1$ is the amplitude, $s=\int c_S d\eta$ is the acoustic
Hubble radius,
  and $\phi$ is the  phase.
 The constants $A_1$ and $\phi$  depend on initial conditions.

We note that baryon pressure has been neglected, $p_b=0$, in the
baryon Euler equation (\ref{velocity-baryon}). Consequently Eq.
(\ref{thetafin}) also assumes that the baryon pressure vanishes.
We discuss this assumption in the following subsection C. Here we
assume vanishing baryon pressure and study acoustic oscillations
driven by a weak Lorentz force. We again neglect gravitational
potentials but now retain the last, Lorentz force, term on the
r.h.s.\ of Eq.~(\ref{thetafin}). In this case the leading JWKB
terms in the  equation  are (again under the assumption that
  ${\dot m}_{\rm eff}/m_{\rm eff} \ll \omega$),
\begin{equation}
{\ddot \Theta_0^{(S)} } + k^2 c_S^2  \Theta_0^{(S)}  = \frac{k^2
L^{(S)}}{4 \rho_\gamma m_{\rm{eff}}}= \frac{3}{4} c_S^2 k^2
\frac{L^{(S)}}{\rho_\gamma}. \label{thetafin11}
\end{equation}
The solutions of Eq.~(\ref{thetafin11}) are of a  similar
 oscillatory
 form to those in Eqs. (\ref{theta00}) and (\ref{theta11}), but now there is a constant
shift of $\Theta^{(S)}_0 \rightarrow \Theta^{(S)}_0 +
3\rho_B/(4\rho_\gamma)$ (here we have used $L^{(S)} \simeq
-\rho_B$), while  $\Theta^{(S)}_1 = - 3  {\dot \Theta}^{(S)}_0/k$
remains unchanged.

\subsection{Acoustic oscillations in the baryon fluid}

The propagation of magnetosonic waves and magnetohydrodynamical
instabilities in the expanding Universe are discussed in detail in
Ref.~\cite{gfd95}. The effects on CMB temperature anisotropies of
magnetosonic waves in a homogeneous magnetic field are studied in
Ref. \cite{Adams}, where  it is shown that a homogeneous magnetic
field induces three types  of MHD waves (fast and slow
magnetosonic and Alfv\'en waves) in an expanding Universe.

Fast magnetosonic waves result in a rescaling   of the fluid sound
speed, i.e., $c_S \rightarrow \sqrt{c_S^2 + v_A^2}$, where the
original fluid sound speed $c_S$ is characterized by the fluid
pressure $p$ and energy density $\rho$, the Alfv\'en speed  $v_A =
B_0/\sqrt{4\pi (\rho+p)}$, and $B_0$ is the (unperturbed
background) homogeneous magnetic field strength \cite{krall}. Fast
magnetosonic waves require a small inhomogeneous magnetic field
$B_1$, so we write the total magnetic field  ${\bf B} = {\bf B}_0
+ {\bf B}_1$, where $|{\bf B}_1| \ll |{\bf B}_0|$. The induction
law in this case is
\begin{equation}
\frac{\partial {\bf B}_1}{\partial t} = {\bf \nabla} \times [ {\bf
v} \times {\bf B}_0 ], \label{bo}
\end{equation}
where we work the leading order in the inhomogeneity. For the
case of  a homogeneous magnetic field, Adams et al. \cite{Adams}
consider  a zeroth-order (background) magnetic field ${\bf B}_0$
with zeroth-order energy density, $\rho_B \propto B_0^2 $, small
 compared to the energy density of the photon-baryon fluid. In this
case ${\bf B}_1$ is a first order perturbation. The fluid
$3$-velocity perturbation ${\bf v}$ is also first  order and
satisfies the linearized Euler equation, \cite{gfd95,Adams},
\begin{equation}
\rho \frac{\partial {\bf v}}{\partial t} + \nabla p +
\frac{1}{4\pi} [{\bf B}_0 \times ( {\bf \nabla} \times {\bf
B}_1)]=0.  \label{VV}
\end{equation}
Here the pressure gradient $ \nabla p$ is related to the sound
speed through $ \nabla p = c_S^2 \nabla \rho$. In addition, the
perturbed magnetic field  obeys the Gauss law ${\bf \nabla} \cdot
{\bf B}_1=0$.

Assuming tight coupling between photons and baryons, multiplying
the baryon Euler  equation by $R$ and adding the photon Euler
equation{\footnote{ The baryon and photon Euler equations are
Eqs. (11) and (13) of Ref. \cite{Adams}.}}, Adams et al.
\cite{Adams} obtain the Euler equation for the photon-baryon
fluid accounting for a zeroth-order spatially homogeneous
background magnetic field (also see Eq. (52) of Ref.
\cite{subramanian05}),
\begin{equation}
\frac{\partial}{\partial \eta}\left[(1+R) v \right] - c_{S,b}^2 k
\delta_b  = \frac{k}{4} \delta_\gamma  + (1+R)k\Psi +
\frac{k}{4\pi (\rho_\gamma + p_\gamma)} [{\bf\hat k} \cdot \{{\bf
B}_0 \times ({\bf\hat k} \times {\bf B}_1)\}], \label{vdot2}
\end{equation}
where $c_{S, b}$ is the baryon fluid (not the photon-baryon
fluid) sound speed.

Compared to the stochastic magnetic field case of
Eq.~(\ref{vdot}), Eq. (\ref{vdot2}) --- for a homogeneous
background magnetic field ---  contains an additional baryon
pressure term on the l.h.s.,\ $c_{S,b}^2 k \delta_b$, while the
term on the r.h.s.\ of Eq. (\ref{vdot2}) $\propto {\hat {\bf k}}
\cdot \{{\bf B}_0 \times ({\bf\hat k} \times {\bf B}_1)\} $ is
the analog of the Lorentz force term $\propto L^{(S)}$ on the
r.h.s.\ of Eq. (\ref{vdot}). We emphasize that Eq.~(\ref{vdot})
is valid for a stochastic magnetic field and that here  the
smoothed amplitude of the magnetic field $B_\lambda$ is $1/2$
order in the perturbation expansion, while ${\bar \rho}_B (\propto
B_\lambda^2)$ and $L^{(S)}$ are first order. We argue below  that
at linear order the additional baryon pressure term ($c_{S,b}^2 k
\delta_b$) in Eq. (\ref{vdot2}) can be neglected.  However,  we
emphasize that even if the sound speed in the uncoupled baryon
fluid vanishes, i.e., $c_{S, b} =0$, the effective sound speed
in  the coupled photon-baryon fluid is not the same as the sound
speed in the uncoupled photon fluid, Due to the tight coupling
between photons and baryons the photon-baryon fluid sound speed
depends on the baryon fraction, $c_S = 1/\sqrt{3m_{\rm eff}}$
(see Eq. (\ref{thetafin111})), so the coupling between baryons
 and photons reduces the sound speed from the $1/\sqrt{3}$
  value for an uncoupled photon fluid.

The last term  on the r.h.s.\ of Eq. (\ref{vdot2}) induces fast
magnetosonic waves in  the presence of a homogeneous magnetic
field. These magnetosonic waves change the photon-baryon fluid
sound speed, increasing it relative to the case without a magnetic
field. In the limit of a weak magnetic field, the effective sound
speed is ${\bar c}_S = \sqrt{1/{(3m_{\rm eff}) + v_A^2}}$
\cite{Adams}. The Alfv\'en speed $v_A$ here is that defined in the
photon-baryon fluid, i.e.,
\begin{equation}
v_A^2 = \frac{B_0^2}{4 \pi (\rho_B + 4\rho_\gamma/3)} =
  \frac{3B_0^2}{16 \pi (1+R) \rho_\gamma}.
\label{alfven-vel}
\end{equation}

For the more realistic case of a stochastic magnetic field, the
Alfv\'en speed should be defined  in the terms of the smoothed
magnetic field $B_\lambda$ \cite{kr05}. Equation (\ref{vdot}) is
the analog of Eq. (\ref{vdot2}) for the case of a stochastic
magnetic field. For this stochastic field case we define the
Alfv\'en speed squared as ${\bar v}_A^2 = 3{\bar \rho}_B/(2(1+R)
\rho_\gamma)$. Here we neglect the baryon pressure, $p_b
\approx0$. Note that for the case $n_B =-3$ our definition of the
Alfv\'en speed coincides with that used in Ref.~\cite{Adams} under
the assumption that $B_\lambda = B_0$, i.e., $v_A={\bar v}_A$.
The term $\propto L^{(S)}$ on the r.h.s.\ of Eq. (\ref{vdot})
ensures that the effective sound speed  is rescaled in a manner
similar to that for an homogeneous magnetic field, $c_S
\rightarrow {\bar c}_S = \sqrt{1/(3m_{\rm eff}) + {\bar v}_A^2}$.

This rescaling of the sound speed may be formally described as a
baryon energy density fraction change $R \rightarrow {\bar
R}=R-\Delta R$, as follows. We define ${\bar m}_{\rm eff} = {\bar
c}_{S}^2 /3$, then
\begin{equation}
{\bar m}_{\rm eff}= \frac{m_{\rm
eff}}{1+9B_0^2/(16\pi\rho_\gamma)},
\end{equation}
so the rescaling of the sound speed induced by the presence of a
magnetic field (a homogeneous or stochastic field) is equivalent
to a reduction of the baryon fraction, $\Delta R= 3{\bar
v}_A^2m^2_{\rm eff}/(1+ 3{\bar v}_A^2m_{\rm eff})$. Consequently,
this increase of the sound speed (relative to that
  of a model without a
magnetic field) induces shifts of the CMB anisotropy angular power
spectrum  peaks comparable to shifts resulting from a reduction of
the baryon density $\rho_b \rightarrow \rho_b - 3(1+R)B_0^2/(4\pi)
$; here we have used the fact that ${\bar v}_A \ll {\bar c}_S$.
This was noted, for a homogeneous magnetic field, from the result
of numerical simulations, in Ref. \cite{Adams}; the  analytical
results we have derived, for a homogeneous or a stochastic
magnetic field, are new. The reduction of the baryon fraction
reduces the baryon drag force ($\propto R k^2 \Psi$) on the
r.h.s. of Eq. (\ref{thetafin}). As a result there are two
effects: a shift of the CMB anisotropy angular power spectrum
peak positions and a reduction of the peak amplitudes.

We also note that the Lorentz force in the Poisson equation
 (\ref{phi1}) can be treated as reducing the photon
effective temperature. In particular, photon energy density
perturbations ($\delta_\gamma$) can be compensated by the Lorentz
force $L^S$ if  $\delta_\gamma (\eta) = 3L^{(S)}/\rho_\gamma $
(also see Eq. (\ref{thetafin11})). Note that
$L^{(S)}/\rho_\gamma$ is time independent.

Ref. \cite{y05} (their Eq. (46)) considers a modified form of Eq.
(\ref{vdot2}) for the case of a stochastic magnetic field: they
discard the Lorentz force term. Ref. \cite{koh} also neglects the
Lorentz force contribution (their Eq. (13) and App. A). Both
references argue that such a force-free approximation is
justified by the infinite conductivity of the plasma which
results in a vanishing electric field in the metric perturbation
equation. We have noted however, at the end of Sec. II, that the
force-free approximation cannot be used in this manner. Here we
point out that the current, $\propto {\bf \nabla \times B}$, does
not vanish, and so a Lorentz force term must be present in the
baryon Euler equation.

We have shown that the Lorentz force term in the baryon Euler
equation results in shifts of the CMB anisotropy angular power
spectrum peaks, relative to the case without a magnetic
field.\footnote{Below we discuss the peak shifts related to the
baryonic pressure effect, which are small compared the shifts
discussed here.} This result, obtained for a stochastic magnetic
field, generalizes that of Ref. \cite{Adams} for an homogeneous
magnetic field; for $n_B=-3$ our estimate reproduces the acoustic
oscillations shown in Fig. 1 of Ref. \cite{Adams}. On the other
hand this result contradicts Fig. 1 of Ref.~\cite{koh}. This
figure  is for the case of a stochastic magnetic field and it
indicates that the magnetic field effectively increases the baryon
fraction (while Ref. \cite{koh} considers a magnetic field with
positive spectral index, $n_B>0$, this cannot explain the
effective increase of the baryon fraction they find).

We note that Ref. \cite{y05} retains the baryon pressure term
$\propto c_{S, b}^2 k \delta_b$  (and argues that  $c_{S,b}^2 $ is
related to the gradient of the magnetic pressure, $\nabla p_B$) in
the photon-baryon Euler equation (\ref{vdot2}). Ref. \cite{y05}
claims that their CMB anisotropy angular power spectrum peak
shifts are similar to those found in Ref. \cite{Adams}, because of
this term.  The $c_{S, b}^2 k \delta_b$ term does induce a
rescaling of the sound speed. However,  this term is second order
($\delta_b$  and $\nabla p_B$ are first order), does not
contribute at linear order, and so can be discarded compared to
the Lorentz force contribution which they neglect.

\subsection{Initial conditions}
The presence of a cosmological magnetic field modifies the initial
conditions for the monopole $\Theta_0^{(S)}$ and gravitational
potential, $\Phi $ and $\Psi$, perturbations. A proper treatment
in an inflation model requires analysis of quantum mechanical
fluctuations during inflation, see, e.g., Refs.\ \cite{ratra88}
for the case without a magnetic field. Here we adopt a more
phenomenological approach.

 As discussed in Refs.~\cite{giovannini04,giovannini05}, there are  three different types
 of perturbations to consider, depending on
 conditions in the radiation dominated epoch:
  (i) adiabatic, where the
initial gravitational potentials are large compared to the
magnetic field fraction of the energy density, i.e., $\rho_B/
\rho_\gamma \ll \Psi_{\rm{in}}$ and $\rho_B/ \rho_\gamma \ll
\Phi_{\rm{in}}$, so the magnetic field energy density may be
ignored;
 (ii) quasi-adiabatic,  where
 $\rho_B/
\rho_\gamma \leq  \Psi_{\rm{in}}$,
 $\rho_B/
\rho_\gamma \leq  \Phi_{\rm{in}}$; and, (iii) isocurvature, where
the magnetic field energy density fraction dominates over the
initial gravitational potentials,  $\rho_B/ \rho_\gamma \gg
\Psi_{\rm{in}}$ and $\rho_B/ \rho_\gamma \gg \Phi_{\rm{in}}$.

Since adiabatic and quasi-adiabatic CMB perturbations have been
studied in some detail (for a review see Ref.
\cite{subramanian05}), we focus on isocurvature fluctuations
induced by a cosmological magnetic field.

 We assume (as is conventional  in the case of an
isocurvature solution) that initial values of all relevant
variables are determined by the magnetic field energy density and
anisotropic stress. Under such an assumption the initial
conditions for the gravitational potentials are obtained through
Eqs.\ (\ref{phi}) and (\ref{psi1+phi1}) (assuming that initial
fluid perturbations are zero),
\begin{eqnarray}
&& k^2 \Phi_{\rm{in}} = \frac{4\pi G}{a_{\rm{in}}^2}
\rho_B(\eta_0),  \nonumber
\\
&& k^2(\Phi_{\rm{in}}+\Psi_{\rm{in}})= -\frac{12\pi
G}{a_{\rm{in}}^2} \Pi^{(S)}(\eta_0), \label{magnetic-initial}
\end{eqnarray}
where $a_{\rm{in}}$ is the value of the scale factor when the
initial conditions are applied. Adding Eqs.
(\ref{magnetic-initial}) and using Eq. (\ref{relation}) implies
that $k^2(2 \Phi_{\rm{in}} + \Psi_{\rm{in}}) = -12\pi G
L^{(S)}(\eta_0) /a^2_{\rm{in}}$.

Using these initial conditions for the gravitational potentials
and assuming a weak magnetic field, one may obtain  solutions
 for scalar magnetic
perturbations with wavelengths larger than  the Hubble radius
(i.e., the leading terms in an expansion in $k\eta \ll 1$).

\section{CMB temperature anisotropies}

In this section we compute the CMB temperature anisotropies due to
the presence of a cosmological magnetic field. We assume the
existence of a cosmological magnetic field on scales larger than
Hubble radius, thus we assume  that a magnetic field has been
generated  during  inflation  \cite{ratra92}. Our analysis below
holds for a magnetic field with  spectral index $n_B$ larger than
$-3$.

  CMB temperature fluctuations are caused by scalar
perturbations due to: i) initial intrinsic inhomogeneities on the
last scattering surface; ii) the relativistic Doppler effect due
to the baryon velocity as the photon propagates to the observer;
iii) the difference in the gravitational potential between the
points of photon emission and the observer (the usual Sachs-Wolfe
effect, SW); and, iv) changes in  the gravitational potential as
the photon propagates (the so-called integrated Sachs-Wolfe
effect, ISW) \cite{hu97,HS,subramanian05}.

Using  the total angular momentum formalism  \cite{hu97},
 the angular power spectrum of the CMB temperature
anisotropy measured today is
\begin{eqnarray} C_\ell^{\Theta \Theta (S)}=
{2 \over \pi} \int d k ~k^2 {\Theta_\ell^{(S) \star} (\eta_0, k)
\over {2\ell +1}} {\Theta_\ell^{(S)} (\eta_0, k) \over {2\ell
+1}}. \label{clt}
\end{eqnarray} Here $l$ is the multipole index and
$\Theta_\ell^{(S)} (\eta_0, k)$ is  the $l$-th multipole moment
of the  (scalar-sourced) temperature fluctuation
 and is determined by the integral solution of the
Boltzmann transport equation \cite{hu97},
\begin{eqnarray}
{\Theta_\ell^{(S)}(\eta_0, k) \over 2\ell +1} &=& \int_0^{\eta_0}\!\!\!
\!\!\!d\eta~e^{-\tau} \left [ \left\{\dot \tau (\Theta_0^{(S)} + \Psi) + {\dot
\Psi} - {\dot \Phi}\right\} j_\ell^{(S,0)}(k\eta_0-k\eta) +
\dot \tau
v_b^{(S)} j_\ell^{(S,1)}(k\eta_0-k\eta)+ \dot \tau P^{(S)}
j_\ell^{(S,2)}(k\eta_0-k\eta) \right ].
\nonumber
\\
&{~~~}&
\label{T-int}
\end{eqnarray} Here
 $P^{(S)}=
(\Theta_2^{(S)} - \sqrt{6} E_2^{(S)})/10$ is the anisotropic
(quadrupolar) part of the Compton scattering cross-section ---
which is a source of CMB polarization anisotropies --- where
$\Theta_2^{(S)}$ and $E_2^{(S)}$ are the temperature and
$E$-polarization quadrupole moments{\footnote{The zeroth-order
term in the expansion of the photon Boltzmann transport equation
for $\ell=2$ relates $\Theta_2^{(S)}$ and $E_2^{(S)}$ as
$E_2^{(S)}=-\sqrt{6} \Theta_2^{(S)}/4$, leading to
$P^{(S)}=\Theta_2^{(S)}/4$ \cite{hu97}.}}, and the radial
functions
\begin{eqnarray} j_\ell^{(S,0)}(x)=j_\ell(x), ~~~~~
j_\ell^{(S,1)}(x)=j_\ell^{\prime}(x), ~~~~~ j_\ell^{(S,2)}(x)={1
\over 2} [3 j_\ell^{\prime \prime} +j_\ell(x)], \end{eqnarray}
where $j_\ell$ is the spherical Bessel function and a prime
represents a derivative with respect to  $x$.

Equation (\ref{T-int}) includes the four effects responsible for
the CMB temperature anisotropies  mentioned above. The initial
photon temperature ($\propto P^{(S)}$), SW ($\propto
(\Theta_0^{(S)}+\Psi )$), and baryon velocity Doppler ($\propto
v_b^{(S)}$) effects are present on the last-scattering surface
and so in Eq.\ (\ref{T-int}) they appear with the factor
$e^{-\tau} \dot \tau$; the ISW effect, $\propto (\dot \Psi - \dot
\Phi)$, contributes from decoupling until today and thus  is
suppressed by the factor $e^{-\tau}$.

We are interested in the large-scale CMB temperature anisotropy
 due to a cosmological  magnetic field. The contribution from the
 ISW effect is negligible when compared
to the SW effect.  Also, compared to the SW effect, the Doppler
term $\propto v_b^{(S)}$ plays a secondary role when $1+R
> 1$ \cite{hu97}. The quadrupole term $P^{(S)} \propto k \Theta^{(S)}_1/{\dot \tau}$ (see Eq. (90) of Ref.
\cite{hu97} and the discussion around Eq. (\ref{PS}) in Sec. V
below) is strongly suppressed for $k \eta \ll 1$, and thus we
neglect it on large angular scales. On large angular scales the
largest contribution arises from the ordinary SW effect so we
approximate the temperature integral solution for the scalar
perturbation as
\begin{eqnarray} {\Theta_\ell^{(S)} (\eta_0, k) \over 2\ell +1}
\simeq \int_0^{\eta_0}\!\!\! d\eta ~e^{-\tau} \dot\tau (\Theta_0^{(S)}
+  \Psi) j_\ell(k\eta_0-k\eta). \label{T-int2}
\end{eqnarray} The visibility function $\dot\tau
e^{-\tau}$ is sharply peaked at decoupling, so we use the
approximation, \cite{HS},
\begin{eqnarray} {\Theta_\ell^{(S)}
(\eta_0, k) \over 2\ell +1} \simeq
[\Theta_0^{(S)}(\eta_{\rm{dec}}) +
\Psi(\eta_{\rm{dec}})]j_\ell(k\eta_0-k\eta_{\rm{dec}})\simeq -{1
\over 3} \Phi(\eta_{\rm{dec}})j_\ell(k\eta_0), \label{Tint3}
\end{eqnarray} where $\eta_{\rm{dec}}$ is the value of conformal time at decoupling.
 This is the familiar SW, $\Phi/3$, result.
This expression for the temperature anisotropy multipole moment
depends on $\Phi(\eta_{\rm{dec}})$, so we must solve for
$\Phi(\eta)$.

 The gravitational potential $\Phi (\eta)$ obeys Eq. (\ref{phi}). To solve this equation
 we
decompose $\Phi$ as
\begin{equation}
\Phi (\eta) = \Phi_1 (\eta) + \Phi_2 (\eta), \label{decomposition}
\end{equation}
where the potential $\Phi_1$ is due to the magnetic field energy
density and $\Phi_2$ is related  to the energy density and
velocity perturbations in the fluids (these perturbations in the
fluid are induced by the magnetic field anisotropic stress). From
Eq. (\ref{phi}),  the potentials $\Phi_1(\eta)$ and $\Phi_2(\eta)$
obey
\begin{eqnarray}
k^2 \Phi_1 &=& 4 \pi G a^2 \rho_B, \label{phi1aa}
\\
k^2 \Phi_2 &=& 4 \pi G a^2 \sum_{f}  {\mathcal D}_f,
\label{phi2aa}
\end{eqnarray}
where ${\mathcal D}_f$ is  the gauge-invariant energy density
perturbation in the $f$-th fluid, Eq.\ (\ref{DD}).

Mathematically ${\mathcal D}_f$ should be obtained through the
solutions for the energy density ($\delta_f$) and velocity
($v_f^{(S)}$) perturbations of the $f$-th fluid. $\delta_f$ and
$v_f^{(S)}$ obey fairly complicated second order differential
equations (see Sec. III.B) that are not straightforward to
integrate. However, since we only consider perturbations arising
from a magnetic field, it is expected that on scales larger than
the Hubble radius the sum of
 the density and velocity perturbations in the fluids,  $\sum_f {\mathcal
 D}_f$,
should be of order the magnetic field energy density, while on
small scales radiation pressure prevents perturbations from
growing \cite{mkk02}. Thus $\sum_f {\mathcal D}_f \leq
\rho_B$.{\footnote{We justify this approximation by using the
fact that the evolution of the magnetic-field-induced energy
density and velocity perturbations on large scales  must follow
the dynamics of the magnetic source.  On the other hand, the
homogeneous solutions of Eq.\ (\ref{phi}) on small scales (in a
model
 without a magnetic field) are sound waves (oscillating energy density
 perturbations).  To obtain solutions of the inhomogeneous equation
 we must
 integrate the product of the source and the
homogeneous solutions; as a consequence of  the oscillation, the
resulting integral will be negligibly small.}} Using this, Eqs.\
(\ref{phi1aa}) and (\ref{phi2aa}) imply that  the maximal value
for $\Phi$  is
\begin{equation}
\Phi (\eta) =\frac{\Phi(\eta_0)}{a^2(\eta)}= \frac{8 \pi G
\rho_B(\eta_0)}{k^2 a^2}. \label{phi14}
\end{equation}

With this expression for $\Phi(\eta)$, the temperature anisotropy
multipole moment, Eq. (\ref{Tint3}), becomes
\begin{eqnarray}
 {\Theta_\ell^{(S)} (\eta_0, k) \over 2\ell +1} \simeq
-{8 \pi G  \over 3 k^2 a^2_{\rm{dec}}} \rho_B(\eta_0, k
)j_\ell(k\eta_0), \label{Tint4} \end{eqnarray} where
$a_{\rm{dec}}$ is the value of the scale factor at decoupling.
Using this in Eq.\ (\ref{clt}), the CMB temperature anisotropy
angular power spectrum is given by
\begin{eqnarray} C_\ell^{\Theta
\Theta (S)} = {2 \over \pi} \left ( {8 \pi G \over 3
a^2_{\rm{dec}}} \right )^2 \int_0^\infty {dk~
{|\rho_B(\eta_0,k)|^2 \over k^2} j_\ell^2(k \eta_0)}.
\end{eqnarray} Now $|\rho_B|^2 $ is given in Eq. (\ref{pi-sca}),
so in terms of the Bessel function
$J_{\ell+1/2}=\sqrt{2x/\pi}~j_{\ell}(x)$, we find
\begin{eqnarray} C_\ell^{\Theta \Theta (S)}= \frac{2 \pi^2 G^2
B_\lambda^4\lambda^{2n_B+6} } {3 (2n_B +3) \Gamma^2 (n_B/2+3/2)
a^4_{\rm{dec}} \eta_0} \int_{0}^{\infty} dk ~{1 \over k^3} \left
[ k_D^{2n_B+3} + {n_B \over n_B+3} k^{2n_B+3} \right ] J_{\ell
+1/2}^2 (k \eta_0). \label{c1}
\end{eqnarray}

The integral in Eq.\ (\ref{c1}) may be evaluated by using
 Eq.~(6.574.2)
of Ref.~\cite{gr}. For $n_B < - {3 / 2}$, when the magnetic source
is dominated by the term proportional to $k^{2n_B+3}$, we find
\begin{eqnarray} \ell^2 C_\ell^{\Theta \Theta (S)} =
{2^{2n_B+1} n_B \Gamma(-2n_B) \over (n_B+3){\Gamma^2 (
{-n_B+1/2})}} A^{(S)}_{\Theta \Theta} \ell^{2n_B+2}, \label{clt1}
\end{eqnarray}
while for $n_B> - {3 / 2}$,  when the magnetic source is dominated
by $k_D^{2n_B+3} $, we have   \begin{eqnarray} \ell^2
C_\ell^{\Theta \Theta (S)} = \frac{1}{2} (k_D
\eta_0)^{2n_B+3}A^{(S)}_{\Theta \Theta} \ell^{-1}. \label{clt2}
\end{eqnarray} In these expressions
\begin{equation}
A^{(S)}_{\Theta \Theta}=\frac{\pi^2 G^2 B_\lambda^4 \lambda^{2n_B
+ 6}}{3(2n_B+3)\Gamma^2(n_B/2+3/2)a^4_{\rm{dec}}\eta_0^{2n_B+2}}.
\end{equation}
Equations (\ref{clt1}) and (\ref{clt2}) describe the angular power
spectrum of the CMB temperature anisotropy induced by  scalar
magnetic perturbations. The maximum growth rate of the power
spectrum $\ell^2 C_\ell$ with $\ell$ is $\ell^{-1}$. This
 occurs for $n_B> -3/2$.
So, as expected, the CMB scalar temperature fluctuations due to a
stochastic magnetic field are strongly suppressed at small angular
scales. The suppression is stronger for an inflation generated
magnetic field with $n_B \rightarrow -3$ ($\ell^2 C_\ell \propto
\ell^{-4}$).

\section{CMB polarization anisotropy}
In this section we compute the scalar magnetic-field-induced CMB
$E$-polarization anisotropy angular power spectrum. Scalar
(density) perturbations only induce electric type
$E$-polarization anisotropies.

In the total angular momentum formalism \cite{hu97}, the
$E$-polarization anisotropy angular power spectrum is
\begin{eqnarray} C_\ell^{EE (S)}= {2 \over \pi} \int dk~ k^2
{E_\ell^{(S) \star} (\eta_0, k) \over {2\ell +1}} {E_\ell^{(S)}
(\eta_0, k) \over {2\ell +1}}, \label{cle}
\end{eqnarray} where $E_\ell$ is the $l$-th multipole moment of the $E$-polarization anisotropy.
It is determined by the integral solution of the Boltzmann
transport equation \cite{hu97},
\begin{eqnarray} {E_\ell^{(S)} \over 2\ell +1} = - {\sqrt 6}
\int_0^{\eta_0} d\eta ~e^{-\tau} \dot \tau ~P^{(S)}(\eta)
~e_\ell^{(S)}(k \eta_0 -k\eta). \label{E-l}
\end{eqnarray} Here $P^{(S)}$ is defined below Eq. (\ref{T-int}) and
$e_\ell^{(S)}$  are the
$E$-polarization radial functions,
\begin{equation}
e_\ell^{(S)} (x)= {\sqrt {{3 \over 8} {(\ell+2)! \over (\ell
-2)!}}} ~{j_\ell(x) \over x^2}.  \label{e-fun}
\end{equation}

Polarization anisotropies, being connected with the anisotropic
stress, are determined by shear viscosity.  They are generated
during last scattering when the tight coupling is enhanced by the
fast growth of the inverse differential visibility function
$1/{\dot \tau}$ \cite{hu97}. In the lowest order of the $k/{\dot
\tau}$ expansion, the set of Boltzmann equation solutions are
given in Eq.~(90) of Ref.~\cite{hu97}, and for the scalar
perturbation mode,
\begin{equation}
P^{(S)}= \frac{2k}  {9\dot \tau} \Theta_1^{(S)}. \label{PS}
\end{equation}
Here $\Theta_1^{(S)}$ is the dipole moment of the temperature
anisotropy and obeys Eq.~(\ref{theta1}). A similar equation holds
for the case of scalar magnetic perturbations, see the discussion
in Sec. 3 of the last of Refs.~\cite{giovannini06}. In this case,
in addition to fluid shear viscosity, there is also a contribution
from magnetic anisotropic stress $\Pi^{(S)}$. Since we focus only
on magnetic-field-induced effects, we consider $\Pi^{(S)}$ to be
the dominant source of polarization. Then, according to the
technique used in Secs.~IV.B---D of Ref. \cite{hu97}, and using
footnotes 2 and 7, we have
\begin{equation}
P^{(S)}(\eta) = -\frac{k^2 \eta} {3 \sqrt{2}\dot \tau \rho_{\gamma
0 }}\Pi^{(S)} =\frac{\sqrt{2}k^2 \eta} {9 \dot \tau \rho_{\gamma 0
}} \rho_B(\eta_0,k). \label{PS1}
\end{equation}

Using these expressions for $P^{(S)}$ and $e_\ell^{(S)}$,
Eqs.~(\ref{PS1}) and (\ref{e-fun}), in Eq.~(\ref{E-l}), we find
\begin{eqnarray} {E_\ell^{(S)} \over 2\ell +1}  =
-\frac{\sqrt{(\ell+2)(\ell+1)\ell(\ell-1)}\rho_B(\eta_0,k)}{3
\sqrt{2}\rho_{\gamma 0}}  \int_0^{\eta_0} \!\!\!d\eta ~
\frac{\eta e^{-\tau}} {(\eta_0-\eta)^2} j_\ell(k \eta_0 -k\eta).
\label{E-lfin}
\end{eqnarray} Here we use again the fact that the visibility function
$e^{-\tau} \dot \tau$ peaks at decoupling and approximate the
$E$-polarization integral solution as
 \begin{eqnarray}
{E_\ell^{(S)} \over 2\ell +1}  \simeq
-\frac{\sqrt{(\ell+2)(\ell+1)\ell(\ell-1)}
\rho_B(\eta_0,k){\eta_{\rm{dec}}}}{3 \sqrt{2} \rho_{\gamma
0}{\dot\tau}(\eta_{\rm{dec}})\eta_0^2} j_\ell(k\eta_0).
\label{Elfin}
\end{eqnarray}

To obtain the angular power spectrum $C_\ell^{EE (S)}$ we need
${\dot\tau}(\eta_{\rm{dec}})$. From Eq.~(C3) of Ref.~\cite{HS},
and assuming for the current value of the baryon energy density
parameter  $\Omega_b (\eta_0) = \rho_b/\rho_{\rm{cr}}=0.05$  and
that the redshift at decoupling $z_{\rm{dec}}=1100$, we get
\begin{equation}
\dot \tau(z_{\rm{dec}})=8.05  \left ({\dot a \over a} \right
)_{\rm{dec}}.  \label{tauzdec}
\end{equation}
Photon-baryon decoupling occurs during the matter dominated
epoch, when $a(\eta) \propto
 \eta^2$,  so
$({\dot a / a})_{\rm{dec}} = {2 / \eta_{\rm{dec}}}$, and ${\dot
\tau}(\eta_{\rm{dec}}) \simeq {16.1 / \eta_{\rm{dec}}}$.

Using Eqs.~(\ref{Elfin}), (\ref{tauzdec}), and (\ref{pi-sca}),
the CMB $E$-polarization angular power spectrum of Eq. (\ref{cle})
is
\begin{eqnarray} C_\ell^{EE (S)}=
 \frac{(\ell+2)(\ell+1)\ell (\ell-1)B_\lambda^4\lambda^{2n_B+6}~\eta^4_{\rm{dec}}}
 {3 \times 2^{14} (2n_B+3) \Gamma^2
 (n_B/2+3/2)\rho_{\gamma 0}^2 ~\eta_0^5}
\int_{0}^{\infty} d k~ k \left [ k_D^{2n_B+3} + {n_B
\over n_B+3} k^{2n_B+3} \right ] J_{\ell +1/2}^2 (k \eta_0).
\label{clint}
\end{eqnarray}

To evaluate this integral we consider two cases: $n_B <-3/2$, when
the integral is dominated by the second term ($\propto
k^{2n_B+3}$); and, $n_B>-3/2$, when the main contribution to the
integral  comes from the term $\propto k_D^{2n_B+3}$. In the first
case, for $-3 < n_B< -2$, the integral may be exactly evaluated
using Eq.~(6.574.2) of Ref.~\cite{gr}),\footnote{The magnetic
field source is non-zero up to the damping scale $k_D$. Since the
integral is dominated by small wavenumbers  we can  replace the
upper-cut-off scale $k_D$ by $\infty$.}
\begin{eqnarray} \ell^2 C_\ell^{EE (S)}\simeq
 \frac{n_B \Gamma(-n_B-2)}{4\sqrt{\pi} (n_B+3)\Gamma(-n_B-3/2)}
 A^{(S)}_{EE}\ell^{2n_B+10} ~~~~~~~~~~~(-3 <  n_B < -2). \label{cle1}
\end{eqnarray} For $-2 \leq n_B < -3/2$ we may evaluate integral
using the semi-analytical approximation of the  Appendix of
Ref.~\cite{kr05}. For $x \gg 1$  $J_{l+1/2}(x) \simeq
\sqrt{2/(\pi x)} \cos [x - (l+1) \pi/2]$,  Eq.~(9.2.1) of
Ref.~\cite{ab}, and replacing the oscillatory function $\cos^2x$
by its r.m.s. value of $1/2$, we get (see Eq.~(B2) of
Ref.~\cite{kr05}) for $n_B = -2$,
\begin{eqnarray} \ell^2 C_\ell^{EE (S)} = - 2  \ln \left
( {k_D \eta_0 \over \ell} \right ) A^{(S)}_{EE}\ell^{6}
~~~~~~~~~~~(n_B=-2), \label{cle2}
\end{eqnarray}
while for $-2 < n_B < - {3 /2}$,  \begin{eqnarray} \ell^2
C_\ell^{EE (S)} \simeq  \frac{n_B} {2(n_B+3)(n_B+2)} (k_D
\eta_0)^{2n_B+4} A^{(S)}_{EE}\ell^6~~~~~~~~~~~~~(-2< n_B <-3/2).
\label{cle3}
\end{eqnarray} For the case when $n_B>-3/2$, using
  $ x_D^{2n_B +3} \int_0^{x_D} dx~ x~  J_{\ell +1/2}^2 (x)
={x_D/\pi} $, we have
\begin{eqnarray} \ell^2 C_\ell^{EE (S)}
\simeq  \frac{8}{9 \pi} (k_D \eta_0)^{2n_B+4} A^{(S)}_{EE} \ell^6
~~~~~~~~~~~~(n_B>-3/2). \label{cle4}
\end{eqnarray}
In these expressions
\begin{eqnarray}
A^{(S)}_{EE}=\frac{B_\lambda^4\lambda^{2n_B+6}\eta^4_{\rm{dec}}}{3
\times 2^{14}
(2n_B+3)\Gamma^2(n_B/2+3/2)\eta_0^{2n_B+10}\rho_{\gamma 0}^2}.
\end{eqnarray} Contrary to the
temperature anisotropy, the $E$-polarization anisotropy angular
 power spectrum $\ell^2 C_\ell^{EE}$ grows rapidly  with $\ell$
 (the fastest growth rate is $\ell^6$). An $\ell^6$ dependence
for $E$-polarization is also expected in the absence of  a
magnetic field \cite{hu97}.

\section{Temperature---polarization cross correlations}

In this section we compute the scalar magnetic-field-induced CMB
temperature---$E$-polarization cross-correlation anisotropy
angular power spectrum \cite{hu97},
\begin{eqnarray}
C_\ell^{\Theta E (S)}= {2 \over \pi} \int dk ~k^2
{\Theta_\ell^{(S) \star} (\eta_0, k) \over {2\ell +1}}
{E_\ell^{(S)} (\eta_0, k) \over {2\ell +1}}. \label{clee}
\end{eqnarray} Using Eqs. (\ref{pi-sca}), (\ref{Tint4}), (\ref{Elfin}), and
(\ref{tauzdec}), we find
\begin{eqnarray} C_\ell^{\Theta E (S)} \simeq l^2
\frac {\sqrt{2}\pi G B_\lambda^4
\lambda^{2n_B+6}\eta^2_{\rm{dec}}} {3 \times 2^{10}
(2n_B+3)\Gamma^2(n_B/2+3/2)\eta_0^3 a^2_{\rm{dec}} \rho_{\gamma
0}} \int \frac{d k}{k} \left [ k_D^{2n_B+3} + {n_B \over n_B+3}
k^{2n_B+3} \right ] J_{l+1/2}^2 (k\eta_0), \label{clet}
\end{eqnarray}
  under the
approximations discussed in the two previous sections.

Evaluating the integral in Eq. (\ref{clet}), we find,  for $-3 <
n_B < -{3/2}$,
\begin{eqnarray} \ell^2 C_\ell^{\Theta E (S)}\simeq
\frac{n_B\Gamma(-n_B-1) }{\sqrt{\pi}(n_B+3)\Gamma(
{-n_B-1/2})}A^{(S)}_{\Theta E}\ell^{2n_B+6}, \label{clte1}
\end{eqnarray}
while for $n_B > - {3 / 2}$, \begin{eqnarray} \ell^2
C_\ell^{\Theta E (S)} \simeq     (k_D
\eta_0)^{2n_B+3}A^{(S)}_{\Theta E} \ell^3, \label{celt2}
\end{eqnarray} where
\begin{eqnarray} A^{(S)}_{\Theta E}=\frac{\sqrt{2}\pi G B_\lambda^4 \lambda^{2n_B+6} \eta^2_{\rm{dec}}}{3 \times 2^{11} (2n_B+3)
\Gamma^2(n_B/2+3/2) a^2_{\rm{dec}} \eta_0^{2n_B+6} \rho_{\gamma
0}}. \label{clte2}
\end{eqnarray} As in the case of temperature fluctuations,
the temperature---$E$-polarization cross-correlation angular power
spectrum $\ell^2 C_\ell^{\Theta E (S)}$
 has  maximum growth rate $\propto \ell^3$ for $n_B >-3/2$.

\section{Conclusions}
We present a systematic discussion of scalar isocurvature
magnetic perturbations (magnetosonic cosmological waves) in a
Universe with  a stochastic primordial magnetic field. We derive
the complete set of equations that govern the dynamics of
 linear magnetic energy density perturbations.

A stochastic magnetic field shifts the acoustic peaks of the CMB
temperature anisotropy angular power spectrum, acting in a similar
way as a reduction of the baryon fraction. This result extends
the work of Adams et al. \cite{Adams} who studied a homogeneous
magnetic field. The second important effect that comes from a
stochastic magnetic field is a non-zero  anisotropic stress which
generates a CMB $E$-polarization anisotropy.

We obtain approximate analytical expressions for the CMB
anisotropy angular power spectra $C_\ell^{\Theta \Theta (S)}$,
$C_\ell^{EE (S)}$, and $C_\ell^{\Theta E (S)}$.  Numerical values
of these spectra depend on four parameters: the cut-off wavenumber
$k_D$; the smoothing length $\lambda$; the amplitude of the
smoothed magnetic field $B_\lambda$; and, the magnetic field
power spectral index $n_B$.

We find that the scalar CMB temperature anisotropy power spectrum
amplitude $C_\ell^{\Theta\Theta (S)}$ rapidly decreases with
increasing $\ell$, consequently, the magnetic field energy density
contribution to the total CMB temperature anisotropy signal is
suppressed at large  multipole number.  On the other hand, the
contribution from the magnetic field anisotropic stress to the
$E$-polarization anisotropy becomes large  on small angular
scales and should be accounted for when estimating the magnetic
field contribution to the  CMB $E$-polarization anisotropy (it
must be added to the vector CMB $E$-polarization)
\cite{lewis04,kr05}.

Since scalar CMB temperature perturbations induced  by a
stochastic magnetic field do not depend on magnetic helicity,
precise measurements of the CMB temperature anisotropy angular
power spectra peak positions and amplitudes, combined with a
CMB-independent measurement of the baryon fraction  and  CMB
polarization Faraday rotation data, should provide information on
the symmetric part of the magnetic field power spectrum, $P_B$.
This determination of $P_B$ together with future CMB
$B$-polarization anisotropy data (which is sensitive to both
power spectra, $P_B$ and $P_H$), could lead to a constraint on
cosmological magnetic helicity.

\acknowledgments We thank Arthur Kosowsky and Andy Mack for
fruitful discussions and suggestions and  acknowledge helpful
comments from R. Durrer, M. Giovannini, D. Grasso,  and K.
Subramanian. T.K. thanks Rutgers University for hospitality when
 part of this  work was done and acknowledges partial support from
 INTAS grant 06-1000017-9258 and Georgian NSF grant ST06/4-096.
This work is supported by DOE grant DE-FG03-99EP41093.

\end{document}